\documentclass[aps,preprint,preprintnumbers,eqsecnum,superscriptaddress,twocolumn,tightenlines,10pt,fleqn,showkeys,showpacs]{revtex4-1}
\usepackage[justification=raggedright,singlelinecheck=false]{caption}

\usepackage{graphicx}
\usepackage{xcolor}
\usepackage{rotating}
\usepackage{bm,amsmath,amssymb}
\usepackage[utf8]{inputenc}
\usepackage{footmisc}
\usepackage{siunitx}
\usepackage{slashed}

\newcommand{\nb}[1]{\color{blue}}

\usepackage{graphicx}
\graphicspath{ {./images/} }

\usepackage{caption}
\usepackage{subcaption}

\usepackage{comment}

\usepackage[normalem]{ulem}
\usepackage{amsmath}
\usepackage{enumerate}
\usepackage{amsfonts}
\usepackage{epsfig}
\usepackage{mathbbol}

\newcommand\half{{\ensuremath{\frac{1}{2}}}}
\newcommand\p{\ensuremath{\partial}}
\newcommand\pp{\ensuremath{\bm\nabla}}

\newcommand\vev[1]{{\ensuremath{\left\langle{#1}\right\rangle}}}

\newcommand\ket[1]{\ensuremath{\lvert{#1}\rangle}}
\newcommand\bra[1]{\ensuremath{\langle{#1}\rvert}}

\newcommand{\be}{\begin{equation}}
\newcommand{\ee}{\end{equation}}
\newcommand{\bea}{\begin{eqnarray}}
\newcommand{\eea}{\end{eqnarray}}
\newcommand{\bega}{\begin{gather}}
\newcommand{\eega}{\end{gather}}

\newcommand{\bi}{\begin{itemize}}
\newcommand{\ei}{\end{itemize}}
\newcommand{\ben}{\begin{enumerate}}
\newcommand{\een}{\end{enumerate}}
\newcommand{\bca}{\begin{cases}}
\newcommand{\eca}{\end{cases}}
\newcommand{\bln}{\begin{align}}
\newcommand{\eln}{\end{align}}
\newcommand{\bst}{\begin{split}}
\newcommand{\est}{\end{split}}
\def\ie{\begin{equation}\begin{aligned}}
\def\fe{\end{aligned}\end{equation}}
\newcommand{\bma}{\le(\begin{matrix}}
\newcommand{\ema}{\end{matrix}\ri)}
\newcommand{\bwt}{\begin{widetext}}
\newcommand{\ewt}{\end{widetext}}

\newcommand\lam{\lambda}
\newcommand\Lam{\Lambda}

\newcommand\Th{{\Theta}}

\newcommand\ov{\over}
\newcommand\ha{{\half}}

\def\le{\left}
\def\ri{\right}

\newcommand\sD{{\ensuremath{{\mathcal D}}}}
\newcommand\sE{{\ensuremath{{\mathcal E}}}}

\newcommand\sI{{\ensuremath{{\mathcal I}}}}

\newcommand\sH{{\ensuremath{{\mathcal H}}}}

\newcommand\sM{{\ensuremath{{\mathcal M}}}}
\newcommand\sN{{\ensuremath{{\mathcal N}}}}
\newcommand\sO{{\ensuremath{{\mathcal O}}}}

\newcommand\sV{{\mathcal V}}

\newcommand\sR{{\mathcal R}}

\DeclareMathAlphabet{\pazocal}{OMS}{zplm}{m}{n}

\begin{document}

\title{Theory of the Photomolecular Effect}

\author{Michael J. Landry}
\thanks{Corresponding authors. \href{mailto:mjlandry@mit.edu}{mjlandry@mit.edu},  \href{mailto:mjlandry@mit.edu}{gchen2@mit.edu} \href{mailto:mingda@mit.edu}{mingda@mit.edu}}
\affiliation{Quantum Measurement Group, MIT, Cambridge, MA 02139, USA}
\affiliation{Department of Nuclear Science and Engineering, MIT, Cambridge, MA 02139, USA}
\affiliation{Department of Physics, MIT, Cambridge, MA 02139, USA}

\author{Chuliang Fu}
\affiliation{Quantum Measurement Group, MIT, Cambridge, MA 02139, USA}
\affiliation{Department of Nuclear Science and Engineering, MIT, Cambridge, MA 02139, USA}
\author{James H. Zhang}
\affiliation{Department of Mechanical Engineering, Massachusetts Institute of Technology, Cambridge,
MA 02139}
\author{Jiachen Li}
\affiliation{Department of Mechanical Engineering, Massachusetts Institute of Technology, Cambridge,
MA 02139}
\author{Gang Chen}
\thanks{Corresponding authors. \href{mailto:mjlandry@mit.edu}{mjlandry@mit.edu},  \href{mailto:mjlandry@mit.edu}{gchen2@mit.edu} \href{mailto:mingda@mit.edu}{mingda@mit.edu}}
\affiliation{Department of Mechanical Engineering, Massachusetts Institute of Technology, Cambridge,
MA 02139}
\author{Mingda Li}
\thanks{Corresponding authors. \href{mailto:mjlandry@mit.edu}{mjlandry@mit.edu},  \href{mailto:mjlandry@mit.edu}{gchen2@mit.edu} \href{mailto:mingda@mit.edu}{mingda@mit.edu}}
\affiliation{Quantum Measurement Group, MIT, Cambridge, MA 02139, USA}
\affiliation{Department of Nuclear Science and Engineering, MIT, Cambridge, MA 02139, USA}

\bigskip

\begin{abstract}
    It is well-known that water in both liquid and vapor phases exhibits exceptionally weak absorption of light in the visible range. Recent experiments, however, have demonstrated that at the liquid-air interface, absorption in the visible range is drastically increased. This increased absorption results in a rate of evaporation that exceeds the theoretical thermal limit by between two and five times. Curiously, the evaporation rate peaks at green wavelengths of light, while no corresponding absorptance peak has been observed. Experiments suggest that photons can cleave off clusters of water molecules at the surface, but no clear theoretical model has yet been proposed to explain how this is possible. This paper aims to present such a model and explain this surprising and important phenomenon. 
\end{abstract}


\maketitle

\section{Introduction}

Water exhibits minimal absorption of light within the visible spectrum, with an absorption length on the order of 40 meters \cite{Hale1973Water,cooper2018solar}. Consequently, direct solar evaporation of water through thermal processes is inefficient without the incorporation of additional absorbing materials \cite{Ghasemi2014Solar, Wang2014Solar, Tao2018Solar}. Surprisingly, recent studies have shown that polyvinyl alcohol (PVA) hydrogel can achieve evaporation rates that surpass the theoretical thermal evaporation limit \cite{Zhao2018superthermal, tu2023plausible}. Other porous materials have also demonstrated this enhanced evaporation effect \cite{Tian2021superthermal, Guo2020Evap, Wang2020Evap}.

Several hypotheses have been proposed to explain these enhanced evaporation rates. One suggestion is that the latent heat of water is reduced within hydrogels \cite{Zhao2018superthermal}, however, thermodynamic and heat transfer analysis have consistently shown that this hypothesis is problematic \cite{Chen2022hydrogel, tu2023plausible, zhang2024heat}. The most compelling hypothesis, as presented in \cite{tu2023plausible,lv2024photomolecular}, is the photomolecular effect. This effect posits that photons can directly cleave clusters of water molecules from the surface at the liquid-air interface. A comprehensive range of experiments supports this hypothesis \cite{tu2023plausible,lv2024photomolecular, Verma2024photomolecular}.

The photomolecular effect shares notable similarities with the photoelectric effect—first observed by Hertz and later explained by Einstein; however, there are three critical differences: (1) no electronic transition is involved, (2) it occurs in the visible spectrum, where bulk water does not absorb, and (3) one photon can cleave off a cluster of molecules rather than a single electron.

Studies have shown that wetted hydrogels exhibit significantly increased absorption of visible light compared to either water or dry hydrogels alone \cite{tu2023plausible}. These findings suggest a surface effect: if water at the liquid-air interface can absorb more light than bulk water, the exceptionally large surface area of the hydrogel can greatly enhance these surface effects. 

A particularly intriguing observation is that green light is exceptionally efficient at driving evaporation. When plotting evaporation rate versus wavelength, there is a pronounced peak at green light. Surprisingly, however, no corresponding peak in the absorption spectrum has thus far been observed. 

Despite the extensive experimental data supporting the photomolecular effect hypothesis, a theoretical framework explaining its microscopic mechanism has been lacking. 
Such a microscopic model could provide key insights into how to increase efficiency for this light-driven evaporation. Around 12\% of industrial energy consumption comes from drying \cite{Kudra2024dry}, and desalination of water has enormous humanitarian benefits \cite{Mekonnen2016water,Nassrullah2020desal}. A theoretical understanding of the photomolecular effect could lead to the creation of hyper-efficient drying technologies and distillation plants. Further, it has been observed that clouds absorb more visible light than simulated results predict~\cite{StephensCloud1990, Cess1995Clouds, Stephens1996Clouds, Crisp1997Clouds}; the photomolecular effect may be the ultimate source of this increased absorption, which could significantly improve the accuracy of climate models. Therefore there is an urgent need to construct a microscopic theory that explains the origins of the photomolecular effect. A recent work proposed a macroscopic model involving generalized Maxwell boundary conditions, demonstrating that the enhanced absorptance can be modeled by Feibleman parameters
\cite{Chen2024Maxwell}, but further work is needed to understand why photomolecular evaporation occurs. This paper aims to provide such a theoretical understanding. We seek to answer several key questions: Why is green light uniquely effective? How can the surface absorption rate lack a green light peak, yet evaporation shows a pronounced peak? How is it that visible light is absorbed at such high rates at the surface when both liquid and gaseous water absorb little visible light? Is this effect inherently quantum mechanical or can it be explained classically?

To address these questions, we begin with a simple mathematical model based on minimal assumptions. This model shows that the peak in the evaporation curve can be explained through a basic counting argument. We then develop a more refined quantum field theoretic model, which elucidates how visible light absorption at the surface can be significantly higher than in the bulk and how a single photon can cleave off an entire cluster of molecules. A crucial yet counter-intuitive feature of our model is that the evaporation peak is not necessarily accompanied by a corresponding absorption peak; that is, green photons need not be absorbed at higher rates than other wavelengths but are in fact able to vaporize water more efficiently. This fundamentally quantum mechanical effect cannot be explained classically and suggests that quantum evaporation can vastly exceed ordinary thermal evaporation. In addition to explaining the available data, we make testable predictions to evaluate the veracity of this theory.

\section{A simple counting argument}

Before building a quantum field theoretic model for the photomolecular effect, it is helpful to understand in principle how such a model might operate. The primary question we aim to address here is: why does one particular wavelength of light (i.e. green light) lead to higher rates of evaporation than other wavelengths in the visible spectrum? One might suppose that this peak in evaporation should result from the resonance absorption of green light by water molecules at the surface. Unfortunately, however, while there exist molecular resonances of water in the infrared and ultraviolet regimes that can lead to significant rates of absorption, no such resonances exist in the visible spectrum \cite{Hale1973Water}. To account for these facts, we employ a model relying on combinatorics to explain the evaporation peak. 

Suppose that an incoming photon strikes the surface of water and scatters one or more molecules into the surrounding air, thereby causing evaporation. If the strength of molecular bonding is $\Delta E$, then a natural assumption is to suppose that a photon can vaporize $n$ or fewer water molecules if its frequency satisfies $\hbar \omega \geq n \Delta E$. Let $p(n,\omega)$ be the probability that a photon of frequency $\omega$ vaporizes $n$ water molecules. Moreover, as a crude approximation suppose that $p(n,\omega) = \Theta(\hbar \omega-n\Delta E) p_n$, that is, so long as the photon has enough energy to vaporize $n$ water molecules, the probability of doing so is frequency-independent. There is no particular reason to suppose this is true; it is just for the sake of simplicity. Later when we consider a more careful microscopic model, we will find deviations from this assumption. 

\begin{figure}
\centering
\includegraphics[width=1\linewidth]{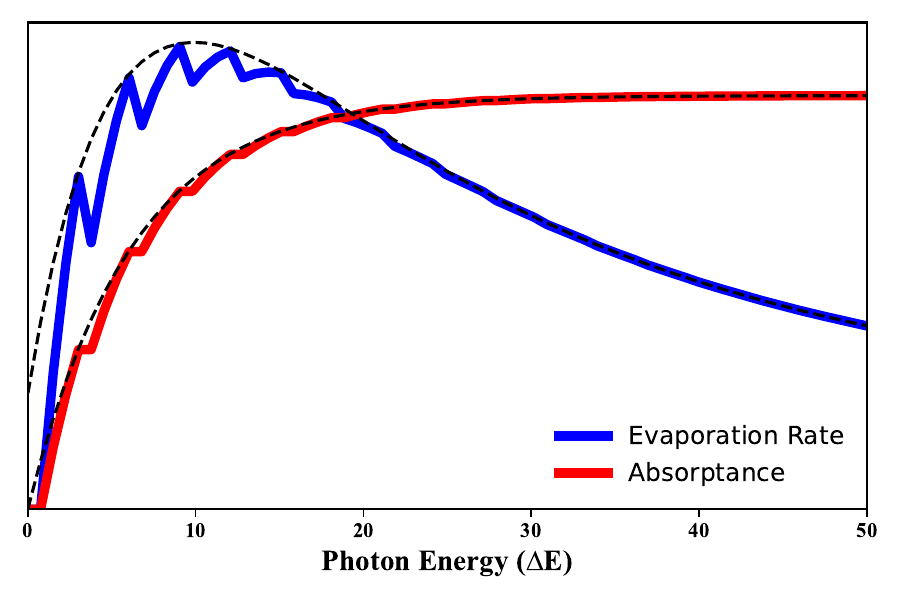}
\caption{depicts the evaporation rate and absorptance, with $\Lam=.85$ chosen for illustrative purposes. Energy is measured in terms of binding energy $\Delta E$. Curves are jagged; dotted lines highlight smooth envelopes. The crucial feature of these plots is that the clear peak in evaporation rate has no corresponding peak in absorptance. }
\label{fig1}
\end{figure}

Consider a scenario in which photons of frequency $\omega$ are shone on the water, and suppose their total amount of energy is $E$. How many water molecules evaporate as a result? If $\omega < \Delta E/\hbar$, clearly nothing will happen. Now suppose frequency $\omega_1 = \Delta E/\hbar$; then a single water molecule can be vaporized per incident photon. 
Moreover, there are $\sN_1 = E/\Delta E$ total photons, so the total number of vaporized water molecules is 
\be N_1 = p_1 E/\Delta E.\ee
Next suppose the photons have frequency $\omega_2 = 2\Delta E/\hbar$. Then the total number of photons is $\sN_2 = E/2\Delta E$ and either one or two water molecules can be vaporized by a single photon. As a result, the number of vaporized water molecules is
\be
     N_2 = \sN_2(p_1+2 p_2) = {E\ov \Delta E}\left(\ha p_1+p_2 \right),
\ee
For the general case of photons of frequency $\omega_n = n\Delta E/\hbar$, the number of vaporized water molecules is
\be
     N_n = {E\ov \Delta E}  \sum_{k=1}^{n} {k\ov n} p_k.
\ee
How should the various probabilities $p_n$ be related to each other? In general, this will be a heavily model-dependent relationship, but we can take a guess. In general, we should expect the likelihood of an $n$-molecule scattering event to get smaller and smaller as $n$ increases, that is $p_{n+1}/p_n <1$. As a very simple model, let us suppose this ratio is independent of $n$ and define $\Lam<1$ by
\be
     \Lam \equiv {p_{n+1}\ov p_n}. 
\ee
Define $p_0 = p_1/\Lam$. Then we find that the total number of vaporized water molecules by photons of frequency $\omega_n$ is
\be\label{H2O}
     N_n = {Ep_0\ov \Delta E} \sum_{k=1}^{n} {k\ov n} \Lam^k,
\ee
and the absorption probability for a given photon is
\be
     \Gamma_n = p_0 \sum_{k=1}^n \Lam^n. 
\ee
We can see in Fig.\ref{fig1} that there exists a peak in the evaporation rate per unit energy, while there is no peak in the absorption rate. Thus, this simple counting argument at least qualitatively explains the data and relies on only three very plausible assumptions:
\begin{itemize}
     \item We assumed the energy required to vaporize $n$ water molecules is simply $n\Delta E$. While this is a reasonable order of magnitude guess, it could be different depending on how bonding works when there are multiple molecules present. 
     \item We assumed that $p(n,\omega)$ has no frequency dependence beyond the cutoff mandated by energy conservation and photon frequency quantization. Explicitly, \be p(n,\omega) = \Theta(\hbar \omega-n\Delta E) p_n.\ee
     \item We supposed that interactions with $n+1$ water molecules is less likely than with $n$; moreover we assumed the ratio $p_{n+1}/p_n =\Lam <1$ is independent of $n$. 
\end{itemize}
When we consider a more careful quantum model, we will see that such assumptions can be slightly relaxed while retaining the important qualitative features.

\section{The quantum model}

The aforementioned mathematical model can explain important qualitative features of the photomolecular effect, but leaves much to be desired. Most notably, we have yet to explain how photons in the visible spectrum are absorbed at the surface at rates that far exceed those of both bulk liquid water and water vapor. Additionally, we might wonder how the angle or polarization of incident light will influence the absorption and evaporation rates, and whether the jagged nature of the evaporation and absorption curves are an artifact of the overly simplistic model or real, experimentally detectable features. Lastly, how does a single photon lead to the vaporization of multiple water molecules?

\subsection{The water cluster size distribution function}

It is well-known that, in the liquid phase, hydrogen bonding leads to the formation of short-lived water clusters or hydrogen-bonded networks \cite{Pettersson2016water,Gao2022cluster,Slaugther1972cluster}. These clusters are bound states of water molecules that exist for brief moments of time; the probability that a given cluster consists of $n$ molecules is given by the water cluster size distribution function $W_n$. We will suppose that a photon, if it possesses sufficient energy, will knock out an entire water cluster at once. Indeed there is experimental evidence to support this assumption \cite{lv2024photomolecular,Verma2024photomolecular}. As a result, we may suppose that the probability $p_n\propto W_n$. 


To model $W_n$, suppose that the energy associated with a cluster consisting of $n$ molecules is given approximately by $G_n = n \Delta G$, for some effective binding energy $\Delta G$. This binding energy represents the average energy connecting a molecule in the cluster to the surrounding water, so a plausible estimate is to suppose $\Delta G \sim 0.01-0.03 {\rm eV}$, namely the van der Waals energy \cite{Vega2005MD,Silvestrelli2009VDW}. Then we may suppose a Boltzmann distribution 
\be W_n\propto e^{-\xi n},\quad \xi\equiv {\Delta G\ov k_B T}.\ee 
Thus, taking $\Lam\equiv e^{-\xi}$, we see that $p_n\propto \Lam^n$, as we supposed in our mathematical model. Such a model, however, neglects the fact that there will be a degeneracy factor for each $n$ such that larger clusters possess higher degeneracy than smaller clusters. We will suppose this degeneracy factor can be modeled by a simple power-law $n^b$ for $b>0$. Comparing with numerical simulations \cite{Gao2022cluster}, we should expect $W_n$ to peak near $n\sim 10$, suggesting that $b=10 \xi$, yielding
\be\label{Wn}
     W_n ={1\ov Z} n^{10 \xi} e^{-\xi n},\quad Z=\sum_{n=1}^\infty W_n. 
\ee
At room temperature, $k_B T\approx 0.03 {\rm eV}$, suggesting that $\xi\sim 0.1-10$.

\begin{figure}
\centering
\includegraphics[width=0.8\linewidth]{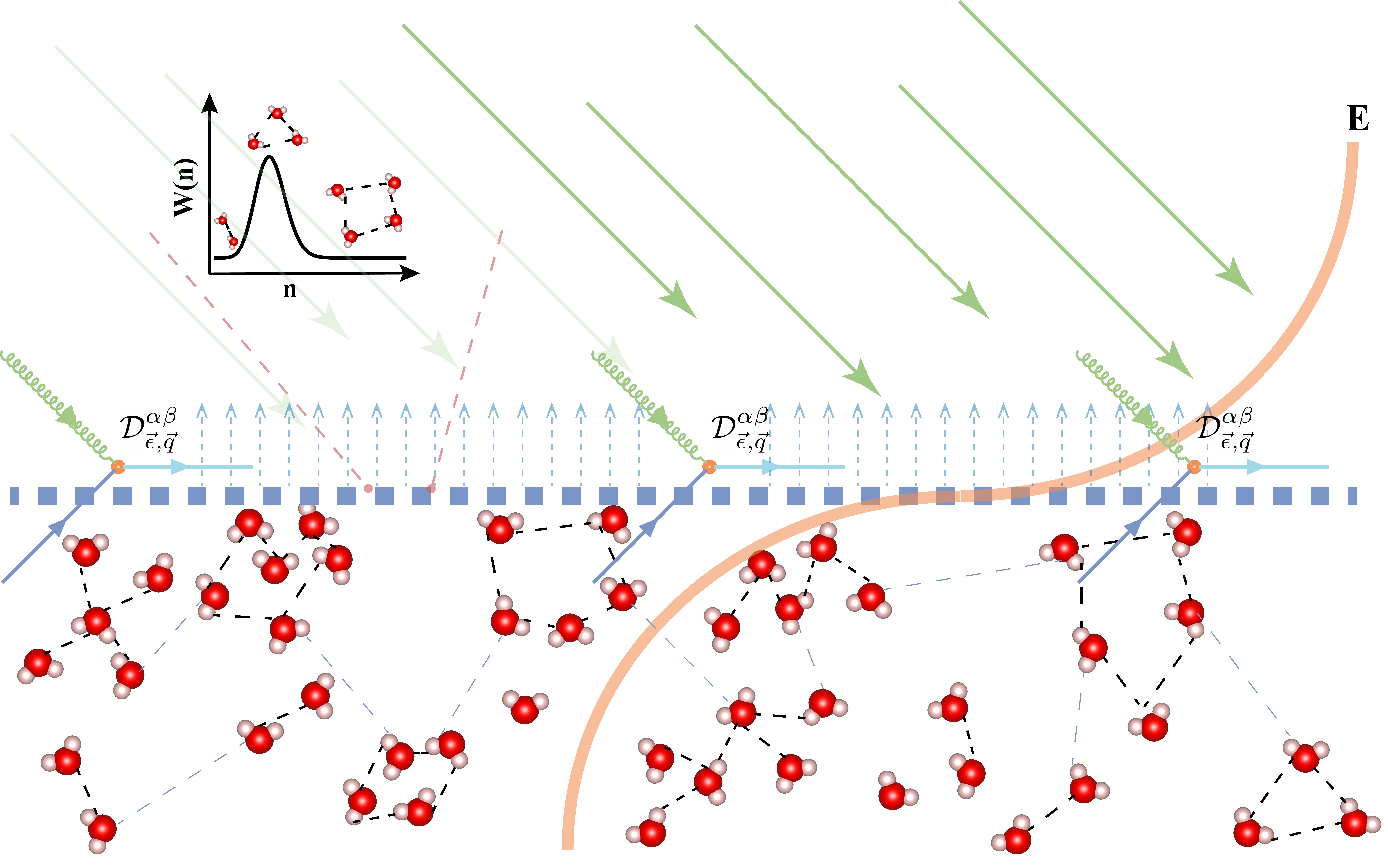}
\caption{Illustrated schematics of the microscopic mechanism behind the photomolecular effect. Sharp electric field gradients of interfacial light generate enhanced cluster-photon interactions with amplitude $\sD^{\alpha\beta}_{\bm \epsilon,\bm q}$, allowing a single photon to vaporize an entire water cluster. The water cluster size distribution function $W_n$ plays an important role in generating the evaporation peak at green wavelengths.}
\label{fig0}
\end{figure}

\subsection{Light-water-cluster interaction}

In order to give a more complete microscopic account, we employ a combination of first and second quantization. The idea is to construct a second-quantized theory describing the interaction of light and water clusters in which the various coefficients are determined using a first-quantized Hamiltonian for the water clusters. 

Consider the structure of the second-quantized Hamiltonian. Let $a^\dagger_{\bm \epsilon,\bm q}$ and $a_{\bm \epsilon,\bm q}$ be the creation and annihilation operators for photons of momentum $\hbar \bm q$ and polarization $\bm\epsilon_{\bm q}$, and let $d_\alpha^\dagger$ and $d_\alpha$ create and annihilate water clusters in quantum state $\alpha$. For the time being, we will consider a single water cluster consisting of $n$ molecules and let the quantum numbers for the water cluster $\alpha,\beta,\dots$ remain generic; we will give a physical account of these quantum states shorty. The second-quantized Hamiltonian can be expressed as 
\bega
\sH = \sH_{\rm EM} + \sH_{\rm cluster} + \sH_{\rm int}^{(1)} + \sH_{\rm int}^{(2)},\\
\sH_{\rm EM} = \sum_{\bm \epsilon\bm q} \hbar \omega_{\bm q} a_{\bm\epsilon,\bm q}^\dagger a_{\bm\epsilon,\bm q},\quad \sH_{\rm cluster} = \sum_\alpha E_\alpha d_\alpha^\dagger d_\alpha,\\
H_{\rm int}^{(1)} = \sum_{\alpha\beta\bm\epsilon\bm q} \sD^{\alpha\beta}_{\bm\epsilon,\bm q} d_\alpha^\dagger d_\beta a_{\bm \epsilon,\bm q} +{\rm h.c.},\\
\begin{split}
H_{\rm int}^{(2)} = \sum_{\alpha\beta\bm\epsilon\bm \epsilon'\bm q\bm q'} \sM^{\alpha\beta}_{\bm \epsilon,\bm \epsilon',\bm q,\bm q'} d^\dagger_\alpha d_\beta a^\dagger_{\bm \epsilon,\bm q} a_{\bm \epsilon',\bm q'} \\
+ \text{two-photon absorption/emission terms. }
\end{split}
\end{gather}
In the above, $\omega_{\bm q} = c q$, where $c$ is the speed of light. The other coefficients $E_\alpha$, $\sD^{\alpha\beta}_{\bm\epsilon,\bm q}$ and $\sM^{\alpha\beta}_{\bm\epsilon,\bm \epsilon',\bm q,\bm q'}$ must be computed for a particular quantum system and can be determined using a first-quantized Hamiltonian in three dimensions. For the moment, treat the cluster as a featureless point-particle (later we will incorporate important effects of its internal degrees of freedom) so the first-quantized Hamiltonian is 
\be
     H = -{\hbar^2\ov 2M} \nabla^2 + V(\bm x),
\ee
where $M$ is the mass of the cluster and $V(\bm x)$ is a potential modeling of the binding of the cluster to the bulk of the water. The resulting energy eigenvalues are $E_\alpha$. To determine the coefficients describing light-cluster interactions, introduce the electromagnetic vector potential $\bm A(\bm x)$ by replacing 
\be
     \pp\to \pp - {i\ov \hbar} (\bm\mu\cdot \pp)\bm A,
\ee
where $\bm \mu$ is the electric dipole moment of the water cluster. Then there are terms $H^{(1)}$ that are linear and $H^{(2)}$ that are quadratic in the photon fields given by
\be
     H^{(1)} = {i\hbar \mu^i\ov 2M} \{\p_j,\p_i A_j\},\quad H^{(2)} = {\mu^i\mu^j\ov 2M} \p_i A_k \p_j A_k.
\ee
The inclusion of $H^{(2)}$ in the first-quantized Hamiltonian and, correspondingly, $\sH^{(2)}_{\rm int}$ in the second-quantized Hamiltonian, is necessary for gauge invariance. We are, however, only interested in the leading-order photon absorption rate, so such terms are irrelevant for present purposes. We will henceforth consider only the single-photon terms $H^{(1)}$ and $\sH^{(1)}_{\rm int}$. 

The atomic vibrational and rotational modes of water molecules have frequencies in the infrared range, while the electronic excitations have frequencies in the ultraviolet \cite{Seki2020waterir,Hayashi2000wateruv}. As the photons are in the visible range, the ultraviolet degrees of freedom can be safely neglected, while the infrared excitations will be significant. We will, however, postpone the treatment of these infrared degrees of freedom, taking $\bm \mu$ constant for the moment.

To account for the fact that the photons exist in the presence of an air-water dielectric interface, we suppose generic mode functions $\bm f_{\bm \epsilon,\bm q}(\bm x)$, resulting in the photon quantum field
\be
     \bm A(t,\bm x) = \sum_{\bm \epsilon\bm q}  \sqrt{\hbar \ov 2 \varepsilon_0 \omega_{\bm q} \sV} \left[e^{-i\omega_{\bm q} t} \bm f_{\bm \epsilon,\bm q}(\bm x) a_{\bm \epsilon,\bm q}+{\rm h.c.}\right],
\ee
where $\varepsilon_0$ is the vacuum permittivity, $\sV$ is the volume of space, $\omega_{\bm q}$ is the frequency of light, and coordinates $(t,\bm x)$ span the entire $3+1$ dimensional spacetime.  In a vacuum, each mode function would take the from $\bm f_{\bm \epsilon,\bm q}(\bm x) = \bm \epsilon_{\bm q} e^{i\bm q\cdot\bm x}$, where $\bm\epsilon_{\bm q}$ is the polarization vector, while in the presence of a dielectric interface at the surface defined by $z=0$, it takes the form\footnote{Here, we only consider in-coming mode functions, that is with $q_z<0$. The out-going mode functions take a similar form after replacing $z\to-z$ and $q_z\to-q_z$.}
\be\label{f}
     \bm f_{\bm \epsilon,\bm q}(\bm x) = \left[\bm \epsilon_{\bm q} e^{i\bm q\cdot\bm x} + \bm R_{\bm\epsilon,\bm q} e^{i\bm q_R\cdot \bm x} \right]\hat\Theta(z) + \bm T_{\bm \epsilon,\bm q} e^{i \bm q_T\cdot \bm x}\hat\Theta(-z), 
\ee
where we take $z>0$ to be air and $z<0$ to be liquid water. We take $\bm q$ and $\bm q_R$ to be, respectively, the incoming and reflected photon wave vector in air, while $\bm q_T$ is the transmitted wave vector in water. Here, $\hat \Th(z)$ is some step-function that looks like the Heaviside theta-function on large scales, but smoothly interpolates between 0 and 1 for $z\sim 0$. We will consider the precise natures of the reflection and transmission coefficients later. 
 This mode function is of crucial importance: the sharp gradient of $\hat \Th(z)$ at the surface of the water will lead to enhanced light-cluster interaction that is absent for water molecules in the bulk or vapor phase. The interaction Hamiltonians in the second quantized theory are specified by those in the first-quantized theory~by
\be
     \sH_{\rm int}^{(i)} \equiv \sum_{\alpha\beta} \bra \alpha H^{(i)}\ket \beta d_\alpha^\dagger d_\beta ,\quad i=1,2,
\ee 
from which the coefficients can be read-off 
\bega
     \sD^{\alpha\beta}_{\bm\epsilon,\bm q} =  \sqrt{\hbar^3\ov8 \varepsilon_0 \omega_{\bm q} \sV} {i\mu^i\ov M} \bra \alpha \{ \p_j,\p_i f^j_{\bm \epsilon,\bm q}(\hat {\bm x}) \} \ket\beta,\\
     \sM^{\alpha\beta}_{\bm\epsilon,\bm \epsilon',\bm q,\bm q'} = {\hbar \mu^i\mu^j\ov 4 M \varepsilon_0  \sV \sqrt{\omega_{\bm q}\omega_{\bm q'}}} \bra\alpha \p_i f^k_{\bm \epsilon,\bm q}(\hat {\bm x}) \p_j f^k_{\bm \epsilon',\bm q'}(\hat {\bm x}) \ket\beta.
\end{gather}

We must now specify an explicit physical model to describe the binding of the water cluster to the surface. We shall suppose a delta-function binding potential $V(\bm x) = - V_0 \delta ^{(3)}(\bm x)$. In this way, there is only one bound state that we represent by $\ket B$. Our model postulates that a photon knocks the cluster out of the bound state $\ket B$ into a scattering state $\ket{\varphi_{\bm k}}$ characterized by momentum $\hbar \bm k$. Letting $E_B$ be the binding energy,\footnote{It turns out that the 3D delta-function potential admits a UV divergence and must be regularized. As such, $V_0$ does not have any independent physical meaning; it is a bare coupling that depends on the UV cutoff. This is OK as all physical quantities are well-behaved; in particular, the physical binding $E_B$ is finite and is specified empirically.} define $\kappa \equiv \sqrt{2 M E_B}/\hbar$; then the wave functions for the scattering and bound states in Fourier space are respectively 
\bega
     \varphi_{\bm k} (\bm p) = (2\pi)^3 \delta^{(3)} (\bm p-\bm k) - {2 g_k\ov p^2-k^2-i0^+}, \\
     \varphi_B(\bm p) ={\sqrt{8\pi\kappa}\ov p^2+\kappa^2},\quad g_k\equiv {2\pi\ov \kappa+ik } .
\end{gather}
See \cite{DeltaFunction1991} (or supplementary material) for a careful derivation of these results. We therefore see that the cluster quantum numbers $\alpha,\beta,\dots$ can take on the value $B$, indicating the bound state, or $\bm k$ representing the scattered (i.e. vaporized) states. 

\subsection{Single-cluster absorption rate}

We now have a complete microscopic model (aside from the internal cluster degrees of freedom) and can compute the absorption rate of a given cluster. To begin let us use Fermi's golden rule to compute the absorption rate of a single photon by a water cluster consisting of $n$ molecules starting off in the bound state $\varphi_B$. To account for the fact that the water cluster can be scattered into the $z>0$ region, but not the $z<0$ region, in which the liquid water resides, we suppose that the only final states the cluster can occupy satisfy $k_z>0$. Then, we may use Fermi's golden rule to compute the single-cluster absorption~rate
\bega
     \Gamma_n(\bm q,\bm \epsilon_{\bm q}) = {2\pi \sN \ov\hbar} \underset {k_z>0} \int {d^3 k\ov (2\pi)^3} \left|\sD^{\bm k,B}_{\bm\epsilon,\bm q}\right|^2\delta(E_{\bm k} - \Delta_{\bm q}),\\
     E_{\bm k} \equiv {\hbar^2k^2\ov 2M},\quad \Delta_{\bm q} \equiv \hbar\omega_{\bm q}-E_B, 
\end{gather}
where $\sN$ is the total number of incoming photons. 
Water clusters are measured in angstroms, while the wavelength of visible light is measured in hundreds of nanometers; as a result we may work in the long-wavelength limit. This limit amounts to neglecting all $\bm x$-dependence in the mode function save for the sharp gradients of $\hat\Th(z)$.  Moreover, the $z$-component of the electric field can be discontinuous at the interface, but the $x$-and $y$-components must be continuous. As a result, only the $z$-component $f^z_{\bm \epsilon,\bm q}(\bm x)$ of the mode function exhibits a sharp gradient. Consequently, the derivative of the mode function is
\be\label{eta}
     \p_i f^j_{\bm\epsilon,\bm q}(\bm x) = \eta_{\bm\epsilon,\bm q} \delta_{zi} \delta_{zj} \hat\delta (z),\quad \eta_{\bm\epsilon,\bm q} \equiv \epsilon_{\bm q}^z + R^z_{\bm\epsilon,\bm q} - T^z_{\bm \epsilon,\bm q}, 
\ee
where $\hat \delta(z)\equiv \p_z \hat\Th(z)$. We should expect the mode function to transition from its vacuum value (in air) to its dielectric value (in water) over length scale $\ell\sim \textup{\AA}$. For small clusters, $E_B$ should be on the order of the van der Waals energy, while for large clusters, it will be close to the photon energy; in every case we find that the cluster wave function has position uncertainty $\kappa^{-1} \ll \ell$. As a result, we may expand the photon mode functions in small $z$ near $z=0$, meaning that $\hat\Th(z)\sim z/\ell $ and $\hat \delta (z) \sim 1/\ell$. Then the single-cluster absorption rate is computed to be (see supplementary material)
\bega
     \Gamma(\bm q,\bm \epsilon_{\bm q}) = {4|\eta_{\bm\epsilon,\bm q}|^2  \mu_z^2\sI\ov 3 \kappa^2\ell^2 \varepsilon_0 c \hbar^2 \omega_{\bm q}} F(x) ,\\ 
     F(x) =  x^{3/2} (1-x)^{3/2}\Th (x) ,\quad x = {1-E_B/\hbar\omega_{\bm q}}, 
\end{gather}

where $\sI \equiv c\hbar \omega_{\bm q} \sN/\sV$ is the intensity of incoming light and $\Th$ is the step function.
Note that $\kappa$ and $E_B$ in the above expression have implicit $n$-dependence, while the cluster electric dipole moment $\mu^z$ should have minimal $n$-dependence \cite{RamiroDipole2006}, which we ignore. We should expect the cluster mass to simply be $M= nm$, where $m$ is the mass of a single water molecule. Similarly, for small clusters, one might suppose that $E_B \propto n \Delta E$, where $\Delta E$ is the binding energy of a single water molecule; this is what we assumed in our simple mathematical model. For large clusters, however, we should suppose that only the water molecules at the surface of the cluster contribute to the binding energy, suggesting a surface area law $E_B \propto n^{2/3}  \Delta E $. To interpolate these two extremes, model $E_B = n^a \Delta E$, where we take $2/3\leq a\leq 1$ to be a fitting parameter. 

While the quantum uncertainty of the cluster's position is much smaller than $\ell$, the physical size of the cluster is larger than $\ell$. If we take the opposite limit and suppose that the mode function is genuinely discontinuous, that is, $\hat\delta(z) = \delta(z)$, then the only change to the single-cluster absorption rate is the mode function must be replaced by (see supplementary material)
\be\label{delta mode}
     F(x) \to {3 \kappa^2\ell^2\ov 16} \sqrt{1-x}\left(\tanh^{-1}\sqrt x-\sqrt x\right) \Th(x). 
\ee
We see, therefore, depending on the precise properties of the mode function $\bm f_{\bm \epsilon,\bm q}(\bm x)$, the form factor $F$ can change. A good model for a generic form factor that can interpolate between these two extremes is 
\be
     F(x) \propto x^{\zeta_1} (1-x)^{\zeta_2} \Th(x). 
\ee
In particular, fixing $\zeta_1=3/2$ and $\zeta_2=1/10$ yields a good approximation of~\eqref{delta mode}. We therefore should expect $1/10\leq \zeta_{1,2} \leq 3/2$.

\subsection{Degeneracy effects}

So far we have computed the absorption rate assuming that the water cluster has no internal structure. In general, this internal structure will be quite complicated and would require intricate numerical computations to fully understand. We can, however, get a sense of its most important feature, namely that it enlarges the final state phase space enormously, especially for larger clusters. This enlarged phase space will vastly increase the absorptance and evaporation rate.

We will make the simplifying assumption that the water cluster remains a single bound state after vaporization. In this case, the internal degrees of freedom can be classified into three categories (1) vibrational modes of each molecule (2) rotational modes of each molecule, and (3) collective modes associated with the relative motion of water molecules within the cluster. Each of these modes has some characteristic length scale $l$; the probability of exciting $\nu$ such modes is suppressed by the factor $(l/\ell)^{2\nu}$ (see supplementary material), where the reader will recall $\ell\sim\textup{\AA}$. The respective characteristic length scales for these three types of modes are
\be
     l_{\rm vib} \sim \sqrt{\hbar\ov m_{\rm H} \omega_{\rm vib}},\quad l_{\rm rot} \sim r_{\rm H_2O} ,\quad l_{\rm coll} \sim r_{\rm clust},
\ee
where $m_{\rm H}=2 \times 10^{-27}{\rm kg}$ is the mass of a hydrogen atom, $\omega_{\rm vib}\sim 4 \times 10^{14} {\rm Hz}$ is the characteristic vibrational frequency of a water molecule, and $r_{\rm H_2O}\sim \textup{\AA}$ and $r_{\rm clust}\sim 1-10\textup{\AA}$ are the radius of a water molecule and cluster, respectively \cite{Gao2022cluster,Santis2024Cluster}. We therefore find that $l_{\rm vib}\sim 0.1 \textup{\AA}$, while $l_{\rm rot}\sim l_{\rm coll} \sim \textup{\AA}$. As a result, excitations of vibrational modes should be suppressed by a factor of $(l_{\rm vib}/\ell)^2\sim 0.01$ and we will hence ignore them. Both rotational modes and collective modes have very little suppression of this variety. The energy cost for exciting rotational modes is the smallest, so we will focus primarily on these internal degrees of freedom. It should be noted, however, that for a more careful treatment of the effective degeneracy factor, collective modes could be quite significant. Indeed the strength of interactions among the molecules of the cluster is considerable, so the distinction between rotational and collective modes could be significantly blurred. 

Suppose that after absorbing the photon and scattering into the final state, $\nu$ internal modes are excited by the photon. If these modes are rotational in nature, letting $I_{\rm H_2O}\sim 2 \times 10^{-46} {\rm kg\, m^2}$ be the moment of inertia for a water molecule, the energy cost will be on the order of $\hbar^2 \nu/I_{\rm H_2O}\sim \nu \times 10^{-3.5} {\rm eV}$, which is negligible even for large $\nu$. Moreover, the interaction energy between molecules comes primarily from hydrogen bonding, which is orders of magnitude larger than the rotational energy. As a result, energy from a single photon can be distributed among rotational degrees of freedom of many molecules. Supposing that each molecule can have at most one quantum of angular momentum, the degeneracy of this process is given by the binomial coefficient $3 n \choose \nu$. 
We will suppose that exciting $\nu$ modes should be more likely than exciting $\nu+1$ modes and denote the suppression factor by $\chi\sim(\ell/l_{\rm rot})^2$. 
As a result, the {\it effective} degeneracy associated with the excitation of $\nu$ internal degrees of freedom is given by 
\be\label{Dn}
     D_n = \sum_\nu {3n\choose \nu} \chi^{-\nu},\quad \chi > 1. 
\ee
This suppression factor's exact value is considered as a fitting parameter. As the length scale $\ell$ is often more precisely estimated to be between $1-3\textup{\AA}$, it is reasonable to choose a value of $\chi$ somewhere in the range $1-10$.

\subsection{Evaporation rate and absorptance}

\begin{figure*}[t]
\centering
\includegraphics[width=1\linewidth]{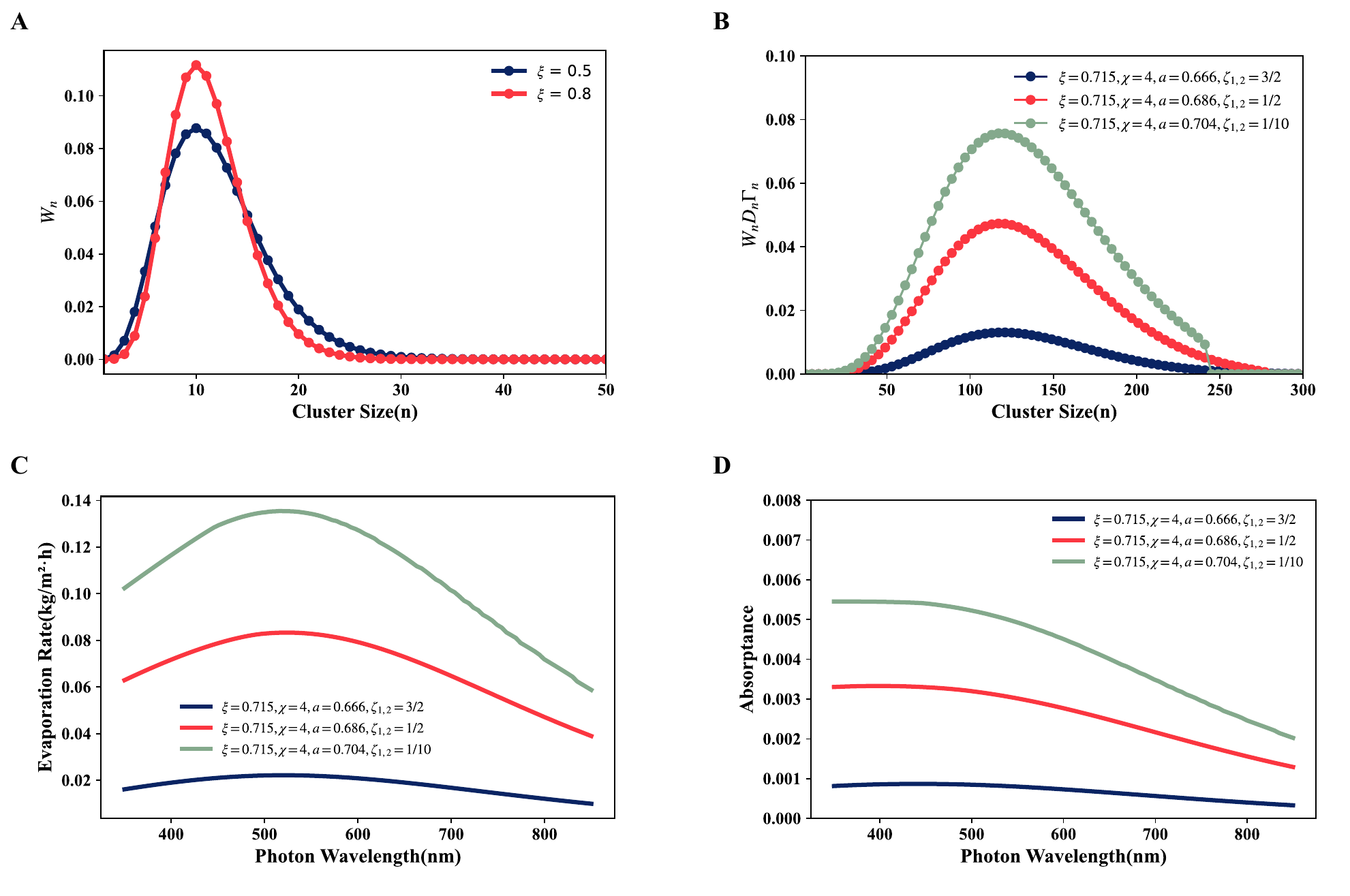}
\caption{The numerical results of the absorption and evaporation. (A) The cluster size distribution function $W_n$, with different $\xi$. The photon wavelength is fixed at 520 nm while the incoming intensity is $1000 \rm{W}/m^2$. It peaked at the cluster size of 10 molecules, the estimated average cluster size. (B) The weighted single cluster absorption rate $W_nD_n \Gamma_n$, with the fixed 520 nm photon wavelength and the incoming intensity is $1000 \rm{W}/m^2$. The degeneracy effect causes the peak shift. (C) The numerical results for the evaporation rate. The different lines represent the different results with different $\zeta_{1,2}$ and modified $a$ to guarantee the evaporation rate peaked at around 520 nm for photon wavelength.  (D) The numerical results for the absorptance. The settings are the same as the case presented in (C). All choices of $\zeta_{1,2}$ yield an evaporation peak at a green light; smaller choices make the peak more pronounced. Crucially, the absorptance curve has no corresponding green peak, which is a characteristic feature of our non-resonant mechanism of absorption and constitutes a clear, testable prediction of our model.}
\label{fig2}
\end{figure*}

With this expression for the single-cluster absorption rate in hand, we can now compute the total absorption rate per cluster, which is given by the weighted average
\be\label{abs2}
     \Gamma_{\rm total}(\bm q,\bm \epsilon_{\bm q}) = \sum_{n=1}^\infty W_n D_n \Gamma_n(\bm q,\bm \epsilon_{\bm q}).
\ee
Letting $\sigma\sim 10^{18} {\rm m^{-2}}$ be the average cluster number-density per unit surface area\footnote{This cluster surface-density can be estimated by noting the averaged number of molecules per cluster is ten and the average spacing between molecules is a few $\textup{\AA}$.} the evaporation rate per unit surface area is
\be\label{evap}
     \sR(\bm q,\bm \epsilon_{\bm q}) = m\sigma \sum_{n=1}^\infty  n W_n D_n \Gamma_n(\bm q,\bm \epsilon_{\bm q}).
\ee
Lastly, we find the absorptance (the ratio of absorbed intensity to incoming intensity measured at the surface) is 
\be\label{totAb}
     \gamma(\bm q,\bm\epsilon_{\bm q}) = {\hbar \omega_{\bm q} \sigma \ov  \sI \cos\theta_I} \Gamma_{\rm total}(\bm q,\bm \epsilon_{\bm q}),
\ee
where $\theta_I$ is the angle of incidence of the incoming photon. 
Fig.\ref{fig2} depicts the evaporation rate and the absorptance. A crucial feature of our model is that there is a peak in the evaporation curve at green wavelengths, but no corresponding green peak in absorptance. We emphasize the fact that this mismatch between evaporation rate and absorptance is only possible in a model in which the evaporation peak does not come from resonance absorption. 
Notice that the jagged curves of~Fig.\ref{fig1} are now absent, meaning they were mere artifacts of the simplistic mathematical model. It should be noted that our model can only be trusted for light in the visible spectrum in which vibrational and electronic forms of absorption are negligible. As a result, the true evaporation and absorption curves should be quite different in the ultraviolet and infrared ranges. 

We determine several related parameters listed below to produce the numerical results presented in Fig.\ref{fig2}. The cluster electric dipole moment is assumed to equal that of a single molecule, namely $\mu^z=1.85$D. The mass of a single water molecule is $m = \num{3e-26} \rm{kg}$. The averaged volume of a water molecule is approximately $\num{3e-29}\rm{nm^3}$, from which we estimate $\sigma = \num{1.036e18} {\rm m^{-2}}$, {assuming there are ten molecules per cluster}. The incoming photon's incidence angle is fixed at $\theta_I = 45^\circ$, with the intensity of incoming light given by one solar intensity, $1000 \rm{W}/m^2$. The single-molecule binding energy is chosen to be $\Delta E = 0.05$ eV, an upper-estimate for van der Waals bond energy.\footnote{It is reasonable to suppose that molecules within the cluster form more than one van der Waals bond with their suroundings.} The length scale is $\ell = \textup{\AA}$. We choose $\xi = 0.715$, $\chi = 4$ to determine the cluster size distribution function and the degeneracy effect. Adjust $\zeta_{1,2} = 3/2, 1/2, 1/10$, and select the corresponding $a = 0.666, 0.686, 0.704$ for the binding energy relation to keep the evaporation rate peaked at around 520 nm, we reach the final simulated results. To get a practical numerical estimation, the summation of the cluster for evaporation rate and absorptance takes the cutoff at $n=300$ since the larger water clusters should have a negligible probability to exist.

\begin{figure*}[t]
    \centering
    \begin{subfigure}{0.9\columnwidth} 
        \centering
        \includegraphics[width=\linewidth]{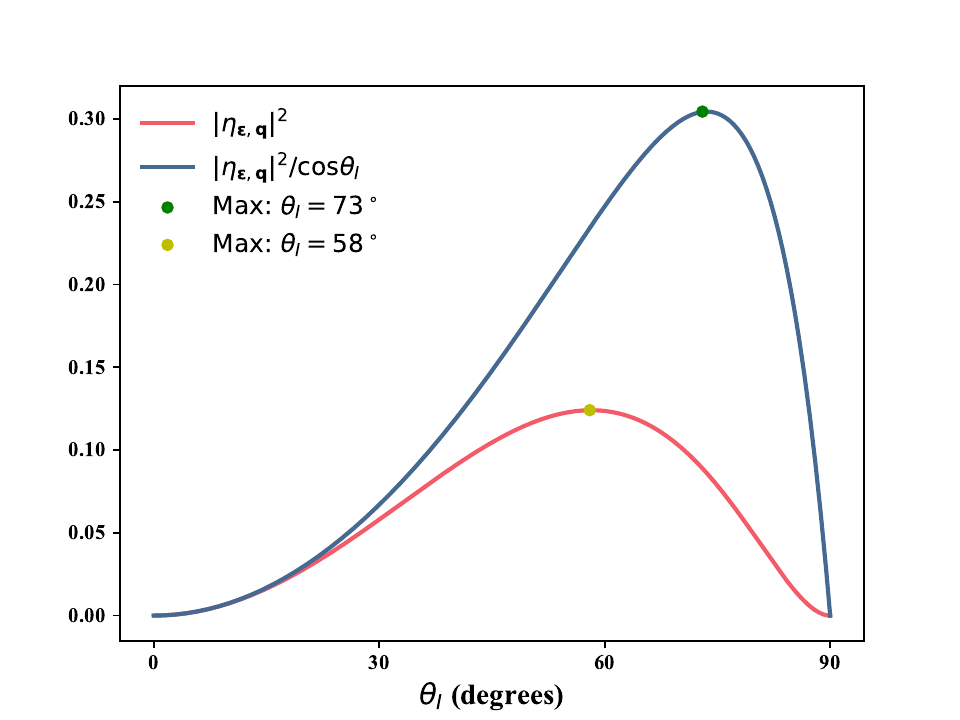} 
        \caption{$\mathbf{A}.$    Light passes from air into water.} 
        \label{fig:subfig1}
    \end{subfigure}
    \hspace{0.02\linewidth} 
    \begin{subfigure}{0.9\columnwidth} 
        \centering
        \includegraphics[width=\linewidth]{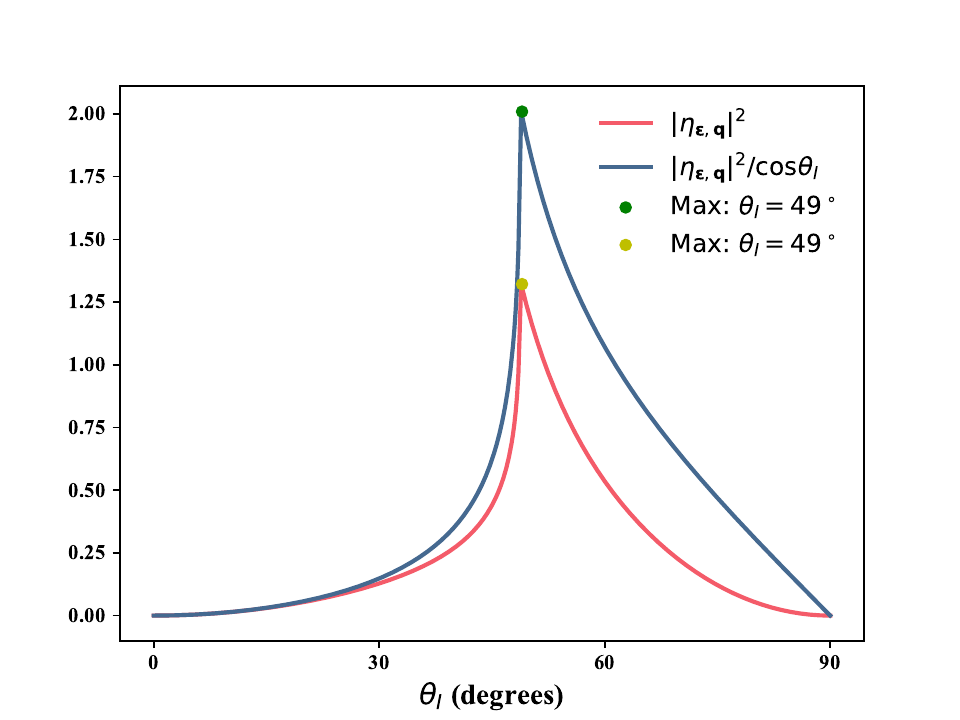} 
        \caption{$\mathbf{B}.$   Light passes from water into air.} 
        \label{fig:subfig2}
    \end{subfigure}
\caption{Angular dependence for the evaporation rate is proportional to $|\eta_{\bm \epsilon,\bm q}|^2$, and for the absorptance is proportional to $|\eta_{\bm\epsilon,\bm q}|^2/\cos\theta_I$. In (A) light passes from air into water so $A=1.33$, while for (B) light passes from water into air so $A=0.75$. Notice a significant increase in absorptance and evaporation rate when light passes from air to water at optimal angles. }
\label{fig3}
\end{figure*}

\subsection{Boundary vs bulk effects}

When devising a model for how water molecules at the boundary can absorb visible light, we must be able to give an account of why this novel mechanism does not lead to the absorption of visible light far from the surface, in the bulk of the water. After all, the absorption of visible light in the bulk is extremely small. There are two key differences between the bulk and the boundary. First, notice that when a water cluster at the boundary absorbs an incoming photon, it can be excited into a free---that is vaporized---state. By contrast, in the bulk no such free states are accessible. Nevertheless, one could imagine a scenario in which a bulk water molecule or cluster might absorb a photon and convert the energy into kinetic motion, which would rapidly dissipate into sound and heat. As such the lack of access to vaporized states, which is characteristic of bulk molecules, is not sufficient to ensure the transparency of liquid water. 

The second, and far more important, difference between the bulk and boundary states involves the mode function~\eqref{f}, which has exceptionally sharp gradients at the surface but small gradients in the bulk. The interaction of light and water that we are positing does not involve excitations that change the magnitude or direction of the intrinsic dipole moment $\bm \mu$. As such this mechanism relies on quadrupole absorption, which typically vanishes in the long-wavelength limit. Thus this effect is extremely small in the bulk, suppressed by a factor of $(\ell/\lam)^2\sim 10^{-7} $,\footnote{This is, in fact, a conservative estimate. The suppression factor could be even more severe, as the degeneracy suppression factor, $\chi$, may be significantly enhanced.} when compared with the absorption process at the surface.

\subsection{Polarization effects}

The effect of the photon's polarization on both absorption and evaporation is captured by $\eta_{\bm\epsilon,\bm q}$. Our present aim is to compute this object. We will suppose that the dielectric coefficient of the surrounding air is $\varepsilon_1$ and that of water is $\varepsilon_2$. Consider $s$-and $p$-polarized light separately. $s$-polarized light is characterized by a polarization vector perpendicular to the plane of incidence, meaning $\epsilon^z_{\bm q} = f_{\bm \epsilon,\bm q}^z=0$ and hence $\eta_{\bm\epsilon,\bm q}=0$. Alternatively, $p$-polarization is characterized by a polarization vector that lies within the plane of incidence, in which case $\epsilon^z_{\bm q} = \sin\theta_I$, where $\theta_I$ is the angle of incidence. Following~\eqref{eta}, we therefore have 
\be
     \eta_{\bm \epsilon,\bm q} = (1+ R(\theta_I) - T(\theta_I)) \sin\theta_I,
\ee
where the reflection and transmission amplitudes are given by the standard results \cite{GriffithsElectro} (see supplementary material),
\bega
     R(\theta_I) = {A^{-1} - B(\theta_I)\ov A^{-1} + B(\theta_I)},\quad T(\theta_I)={2\ov A^{-1} + B(\theta_I)} ,\\
     A = \sqrt{\varepsilon_2\ov \varepsilon_1},\quad B(\theta_I) = {\sqrt{1-A^2 \sin^2\theta_I} \ov\cos\theta_I}. 
\end{gather}
As a result, 
\be
     \eta_{\bm \epsilon,\bm q} =  {2(1-A^{-2}) \ov 1+A^{-1} B} \sin\theta_I. 
\ee
In general, $\varepsilon_{2}$ can be a function of frequency. In the visible spectrum, we should expect $\varepsilon_2\approx 1.77$, while $\varepsilon_1\approx 1$, meaning $A\approx 1.33$ \cite{Hale1973Water}. For a graphical representation, consult Fig.~\ref{fig3}. We see that a maximum in absorptance and evaporation occur respectively at $\theta_I= 73^\circ$ and $\theta_I = 58^\circ$. These results are in good agreement with both experiments \cite{lv2024photomolecular} and the recent macroscopic, phenomenological model \cite{Chen2024Maxwell}. Indeed a direct comparison with this phenomenological model tells us that the imaginary parts of the perpendicular and parallel Feibelman parameters are (see supplementary material)
\be
     d_{\perp i} \sim 17.5 \textup{~\AA},\quad d_{\parallel i} = 0. 
\ee

\section{Discussion}

Our theoretical model offers a comprehensive explanation for the experimental findings that constitute the photomolecular effect. It postulates a novel mechanism whereby visible light---typically minimally absorbed by bulk water---is instead absorbed at the water surface at exceptionally high rates. This increased absorption enables a single photon to vaporize an entire cluster of water molecules, resulting in a fundamentally non-thermal mechanism for evaporation. Remarkably, the entire energy of the absorbed photon is used to vaporize the water cluster, allowing for a hyper-efficient evaporation process in which no energy is lost to heat. This efficiency exceeds that of even the most optimal thermal evaporation methods.

Based on combinatorial arguments, the model provides a non-resonant explanation for the observed evaporation peak at green wavelengths. This leads to the counterintuitive prediction that there should be no corresponding absorptance peak at green wavelengths---a prediction that aligns with current observations, though further experiments are needed to conclusively determine the absence of such a peak. Additionally, the model predicts a polarization dependence for both absorptance and evaporation rates: 
$s$-polarized light should yield negligible absorptance and evaporation, while $p$-polarized light should produce peaks in both. These predictions are in good agreement with current experimental data. A further prediction that has yet to be tested is the location of evaporation and absorptance peaks when light passes from the water to the air. Our model gives clear predictions for the locations of such peaks and predicts that the peak absorptance and evaporation rates should be substantially larger compared with the case in which light passes from air into water.  

While our minimal model successfully explains existing experimental data and makes novel predictions, several refinements could enhance its accuracy and applicability. The most important refinement would be to treat the internal degrees of freedom of the vaporized cluster more carefully. Our minimal model estimates their effect on the final-state phase space, allowing us to parameterize our ignorance. An {\it ab initio} calculation of the internal degrees of freedom could vastly improve the predictive power of this model. As the interplay between the degeneracy factor coming from these internal degrees of freedom and the water cluster size distribution function is of crucial importance, more precise numerical modeling of the large $n$ tail of $W_n$ could vastly enhance the predictive power of this model. Additionally, more refined models for the cluster binding potential $V(\bm x)$ and photon mode functions $\bm f_{\bm \epsilon,\bm q}(\bm x)$, would give a more complete picture. In particular, the surface of water is generally rough on scales of order $\ell\sim{ \textup{\AA}}$, with water molecules moving around chaotically. As such, a more careful treatment would include a statistically averaged mode function to account for this roughness. This roughness would lead to a small, but non-zero, absorptance for $s$-polarized light, which has been experimentally observed \cite{lv2024photomolecular}. 
Additionally, the Van der Waals binding energy is of the order of $k_B T$ (at room temperature), indicating that thermal effects may influence the photomolecular effect and should be included in a more comprehensive model.

The implications of the photomolecular effect are significant for optimizing water evaporation processes. The findings suggest that leveraging this effect could enable the design of materials and systems that maximize evaporation efficiency using specific wavelengths of light. This increased efficiency could be particularly valuable for applications requiring rapid and efficient water removal, such as in water purification technologies, where the photomolecular effect could potentially enable a highly efficient form of distillation.

\acknowledgments{The authors thank K Nelson, Jun Liu and A Maznev for the insightful discussions. M Landry acknowledges support from the National Science Foundation (NSF) Convergence Accelerator Award No. 2235945. CF acknowledges support from DOE Award No. DE-SC0021940. ML is partially supported by the Class of 1947 Career Development Chair and support from R Wachnik. JHZ, JL, and GC are supported by the MIT Boae Award, J-WAFS, and UMRP. JHZ is also supported by the MathWorks Graduate fellowship.}

\bibliography{pnas-sample}

\pagebreak
\widetext
\center

\setcounter{page}{1}

\large
Supplemental Information

\vspace{0.5in}

\textbf{Theory of Photomolecular Effect}

\vspace{0.5in}


\normalsize
\raggedright
This supplementary file contains the following elements:\\
\textbf{Supplementary Text}\\

\makeatletter
\renewcommand \thesection{S\@arabic\c@section}
\renewcommand\thetable{S\@arabic\c@table}
\renewcommand \thefigure{S\@arabic\c@figure}
\renewcommand \theequation{S\@arabic\c@equation}
\setcounter{figure}{0}
\setcounter{equation}{0}
\setcounter{figure}{0}
\setcounter{table}{0}
\setcounter{section}{0}
\makeatother

\section{3D delta function potential}

The following discussion is based on~\cite{DeltaFunction1991}. Consider the water cluster bound to the surface of water. Start with the first quantized Hamiltonian with a $\delta$-function potential in 3D 
\be
     H = -{\hbar^2\ov 2M}\pp^2 + V_0 \delta ^{(3)} (\bm x). 
\ee
Let $\psi(\bm x)$ be an energy eigenfunction with energy $E$ such that 
\be
     \left[E+{\hbar^2\ov 2M }\pp^2\right] \psi = V_0\delta^{(3)}(\bm x) \psi. 
\ee
Define $\sE = E M /\hbar^2$ and $v= V M/\hbar^2$ to obtain 
\be
     \left[\sE + \ha \pp^2 \right]\psi = v\delta^{(3)}(\bm x) \psi. 
\ee
Define the Fourier transform wave function $\varphi(\bm p) = \int d^3 \bm x e^{-i\bm p\cdot \bm x} \psi(\bm x)$; the Schrödinger equation becomes 
\be
     \ha (p^2 - k^2) \varphi(\bm p) = - v\psi(0),\quad \sE = \ha k^2. 
\ee
The scattering solutions are given by the Lippmann-Schwinger equation 
\be
     \varphi_{\bm k}(\bm p) = (2\pi)\delta^{(3)}(\bm p-\bm k) - {2 v \psi(0) \ov p^2 - k^2 - i 0  }. 
\ee
We can self-consistently determine $\psi(0)$ 
\be
     \psi(0) = \int{d^3\bm p\ov(2\pi)^3} \varphi(\bm p) = 1 - 2 v I(- k^2-i 0)\psi(0),\quad I(z) \equiv \int{d^3\bm p\ov (2\pi)^3} {1\ov p^2+z}. 
\ee
It turns out $I(z)$ is divergent, so we must introduce an ultraviolet cutoff $|\bm p| < \Lam$ to find 
\be
     I^\Lam(z) = {\Lam \ov 2\pi^2} - {\sqrt z\ov 4\pi} + \sO(1/\Lam),
\ee
from which we find 
\be
     v \psi(0) = {v \ov 1+ 2vI(-k^2-i0)} = \left( {1\ov v} + {\Lam \ov \pi^2} + {ik \ov 2\pi}\right)^{-1}.
\ee
Introducing the renormalized coupling constant $g$ defined by $1/g \equiv 1/v +\Lam/\pi^2$, gives 
\be
     v \psi(0) = {v \ov 1+ 2vI(-k^2-i0)} = \left( {1\ov g} + {ik \ov 2\pi}\right)^{-1}.
\ee
We then have the scattering wave function  
\be
     \varphi_{\bm k}(\bm p) = (2\pi)^{(3)}(\bm p-\bm k) - {2 g_k\ov p^2-k^2-i 0},\quad g_k \equiv \left({1\ov g} + {ik\ov 2\pi}\right)^{-1}.
\ee

The bound state is given by
\be
     \varphi_B(\bm p) = - {2v \psi_B(0)\ov p^2+\kappa^2},\quad \sE_B = -\ha \kappa^2. 
\ee
The normalization coefficient can be fixed by requiring $\int {d^3 {\bm p}\ov(2\pi)^3} |\varphi_B(\bm p)|^2 =1$, from which we find
\be
     \varphi_B(\bm p) = {\sqrt{8\pi\kappa}\ov p^2+\kappa^2}. 
\ee
This normalization coefficient then specifies a relationship between $\kappa$ and $g$, namely
\be
     \kappa = {2\pi\ov g},\quad g_k = \left( {\kappa+ik\ov 2\pi}  \right)^{-1} =  { 2\pi\ov \kappa+ik}  
\ee

\section{Single-cluster absorption rate}

The rate at which a water cluster in state $\ket{i}$ is scattered to state $\ket f$ by a photon of momentum $\hbar \bm q$ is given by Fermi's Golden rule,
\be
     \Gamma_{i\to f} = {2\pi \ov \hbar} |\vev{f|H_{\rm int}|i}|^2 \delta(\hbar \omega_{\bm q} - (E_f-E_i)).
\ee
Treating the water cluster as a point particle, we take the initial state to be the bound state $\ket{i} = \ket B$, and the final state to be the scattering state $\ket f = \ket{\varphi_{\bm k}}$. Further, let $\sN$ be the total number of incoming photons. We do not care about the final state of the photon, so we will sum over them. We thus have the total absorption rate 
\be
     \Gamma(\bm q,\bm \epsilon_{\bm q}) = {2\pi \sN\ov\hbar} \int_{k_z>0} {d^3 k\ov (2\pi)^3} \left|\vev{\varphi_{\bm k} | H^{(1)} | B}\right|^2 \delta(\hbar \omega_{\bm q} - (E_{\bm k}+E_B)),\quad E_{\bm k} = {\hbar^2 k^2\ov 2M},
\ee
where we suppose $k_z>0$ as the cluster can only scatter into the air and not backward into the bulk. 
Here we take the energy of the bound state to be $-E_B$ so that $E_B$ is a positive quantity. This expression can be rewritten as
\be
     \Gamma(\bm q,\bm \epsilon_q) = {2\pi \sN\ov\hbar} \int_{k_z>0} {d^3 k\ov (2\pi)^3} \left|\sD^{\bm k,B}_{\bm\epsilon,\bm q}\right|^2 \delta(\hbar \omega_{\bm q} - (E_{\bm k}+E_B)),\quad E_{\bm k} = {\hbar^2 k^2\ov 2M}. 
\ee
It is convenient to define 
\be
     \ket{\varphi_{\bm k}} = \ket{\bm k} + \ket{\tilde\varphi_{\bm k} },\quad \tilde\varphi_{\bm k}(\bm p) = - {2 g_k\ov p^2-k^2-i0^+}. 
\ee
To compute $\sD^{\bm k,B}_{\bm\epsilon,\bm q}$, we must consider two cases separately.

\subsection{Gradual mode function}
Suppose that the mode function is linear in $z$ over the length of the cluster. Then we can take $\p_z f^z_{\bm\epsilon,\bm q} = \eta_{\bm\epsilon,\bm q}/\ell$, where $\ell$ is the characteristic lengths scale over which the mode function sharply transitions.  We have 
\bega
     \sD_{\bm \epsilon \bm q}^{\bm k B} = \sqrt{\hbar^3\ov 8\varepsilon_0 \omega_{\bm q} \sV} {i\mu^i\ov M} \vev{\varphi_{\bm k}|\{\p_j , \p_i f_{\bm \epsilon,\bm q}^j\}|B} 
     = \sqrt{\hbar^3\ov 8\varepsilon_0 \omega_q \sV} {i\mu^z\ov M } \eta_{\bm\epsilon,\bm q}{2\ov \ell}(I+\tilde I),
\end{gather}
where $I = \bra{\bm k} \p_z \ket{\varphi_B}=- i k_z \varphi_B(k)$ and $ \tilde I = \bra{\tilde \varphi_{\bm k}} \p_z \ket{\varphi_B}=-\int {d^3 p\ov(2\pi)^3}p_z \tilde \varphi_{\bm k}(\bm p) \varphi_B(\bm p)=0$,
where the last equality follows from the fact that the integrand is odd in $p_z$. Consequently, the absorption rate is 
\bega
     \Gamma(\bm q,\bm \epsilon_{\bm q}) = {2\pi \sN\ov \hbar} {\hbar^3\ov 8\varepsilon_0 \omega_q \sV}{\mu_z^2\ov M^2} |\eta_{\bm\epsilon,\bm q}|^2 {4\ov \ell^2} {2\pi\ov(2\pi)^3} \int dk k^2 I_0(k) \delta(E_{\bm k}-(\hbar\omega_{\bm q}-E_B)),\\
     I_0(k)\equiv \int_0^1 d \cos\theta {8 \pi \kappa k^2_z \ov (k^2+\kappa^2)^2 } = {4\pi \kappa k^2\ov (k^2+\kappa^2)^2},\quad k_z\equiv k\cos\theta.
\end{gather}
Defining the intensity $\sI\equiv c\hbar \omega_{\bm q} \sN/\sV$, 
\bega
     \Gamma(\bm q,\bm \epsilon_{\bm q}) = {4 |\eta_{\bm \epsilon,\bm q}|^2  \mu_z^2\sI\ov 3 \kappa^2 \ell^2 \varepsilon_0 c \hbar^2 \omega_{\bm q}} F(x) ,\quad
     F(x) =  x^{3/2} (1-x)^{3/2}\Th(x) ,\quad x = {1-E_B/\hbar\omega_q}. 
\end{gather}

\subsection{Sharp mode function }
Suppose that the mode function transitions very rapidly from its vacuum value to its dielectric value. Then we may approximate $\hat\Theta(z)=\Th(z)$, which implies $\p_z f^z_{\bm\epsilon,\bm q} = \eta_{\bm\epsilon,\bm q} \delta(z)$ in the long-wavelength limit. We have 
\bega
     \sD_{\bm\epsilon,\bm q}^{\bm k B} = \sqrt{\hbar^3\ov 8\varepsilon_0 \omega_q \sV} {i\mu^i\ov M} \vev{\varphi_{\bm k}|\{\p_j , \p_i f_{\bm\epsilon,\bm q}^j\}|B} 
     = \sqrt{\hbar^3\ov 8\varepsilon_0 \omega_q \sV} {i\mu^z\ov M} \eta_{\bm\epsilon,\bm q}(I_1+\tilde I_1+I_2+\tilde I_2),
\end{gather}
where
\bega
     I_1 = \vev{\bm k| \p_z \circ \delta(\hat z) |B },\quad \tilde I_1 = \vev{\tilde\varphi_{\bm k}|\p_z \circ \delta(\hat z) |B } ,\\
     I_2 = \vev{\bm k| \delta(\hat z)\circ  \p_z |B },\quad \tilde I_2 = \vev{\tilde\varphi_{\bm k}| \delta(\hat z)\circ  \p_z |B }.
\end{gather}
We find that
\bega
     I_1 = \int d^3 \bm x \int {d^3 \bm p\ov (2\pi)^3} \bra{\bm k} \p_z \ket{\bm x} \bra{\bm x} \delta(z) \ket{\bm p}\bra{\bm p} B\rangle \\
     = -ik_z\int {d p_z\ov 2\pi}    {\sqrt{8\pi\kappa}\ov p_z^2 + k_\parallel^2 +\kappa^2} \quad
     = -i k_z \sqrt{2\pi \kappa\ov k_\parallel^2 + \kappa^2},
\end{gather}
while $I_2=\tilde I_1=\tilde I_2=0$, as they all involve integrands that are odd in the momentum $p_z$, which is integrated over. Hence,  the absorption rate is 
\bega
     \Gamma(\bm q,\bm \epsilon_q) = {2\pi\sN\ov \hbar } {\hbar^3\ov 8\varepsilon_0 \omega_{\bm q} \sV} {\mu_z^2\ov M^2} |\eta_{\bm \epsilon,\bm q}|^2 {2\pi\ov(2\pi)^3} \int_0^\infty dk k^2 I_3(k) \delta(E_{\bm k} - (\hbar\omega_{\bm q} - E_B)),\\
     I_3(k) \equiv   \int_0^1 d\cos\theta k_z^2 {2\pi \kappa\ov k_\parallel^2 + \kappa^2},
\end{gather}
which can be evaluated to give 
\bega
     \Gamma(\bm q,\bm \epsilon_q) = {4|\eta_{\bm \epsilon,\bm q}|^2 \mu_z^2\sI\ov 3 \kappa^2 \ell^2  \varepsilon_0 c \hbar^2 \omega_{\bm q}} F(x) ,\\ 
     F(x) = {3 \kappa^2 \ell^2 \ov 16}\sqrt{1-x}(\tanh^{-1}\sqrt x-\sqrt x)\Th(x),\quad x = {1-E_B/\hbar\omega_{\bm q}}. 
\end{gather}

\section{Angular dependence and Feibelman parameters }

Define $\eta_{\bm\epsilon,\bm q} \equiv \epsilon_{\bm q}^z + R_{\bm \epsilon ,\bm q}^z -T_{\bm \epsilon,\bm q}^z$, and note that $\epsilon_{\bm q}^z$, $R_{\bm \epsilon,\bm q}^z$ and $T_{\bm \epsilon,\bm q}^z$ are the $z$-components of the incident, reflected, and transmitted electric fields, respectively, up to overall multiplicative coefficient.  

Let $\varepsilon_1$ and $\varepsilon_2$ be the permittivity of the air and water respectively; the refractive indices are then $n_{1,2}=\sqrt{\varepsilon_{1,2}}$. 
Only $p$-polarized light---light with polarization vector in the plane of incidence---will be considered here. The relevant (dis)continuity equations are 
\be
     \epsilon_1(\epsilon_{\bm q}^z + R_{\bm \epsilon,\bm q}^z) = \epsilon_2 T_{\bm \epsilon,\bm q}^z ,\quad \epsilon_{\bm q}^x + R_{\bm \epsilon,\bm q}^x = T_{\bm \epsilon,\bm q}^x. 
\ee
Here we are taking the plane of incidence to be the $x,z$-plane. Letting $\theta_I$ be the angle of incidence and $\theta_T$ be the angle of transmission, these (dis)continuity equations become
\bega
     \varepsilon_1(-\epsilon_{\bm q} \sin\theta_I + R_{\bm \epsilon,\bm q} \sin\theta_I) = \varepsilon_2(-T_{\bm \epsilon,\bm q} \sin \theta_T),\quad \epsilon_{\bm q} \cos\theta_I + R_{\bm \epsilon,\bm q} \cos\theta_I = T_{\bm \epsilon,\bm q} \cos\theta_T \\ \implies
     \epsilon_{\bm q}  =  R_{\bm \epsilon,\bm q}  + {\varepsilon_2\ov \varepsilon_1}  {\sin\theta_T\ov\sin\theta_I} T_{\bm \epsilon,\bm q},\quad \epsilon_{\bm q}  = - R_{\bm \epsilon,\bm q} + T_{\bm \epsilon,\bm q}{ \cos\theta_T\ov\cos\theta_I} .
\end{gather}
According to Snell's law, ${\sin\theta_T / \sin\theta_I}= \sqrt{\varepsilon_1/\varepsilon_2}$. Consequently,
\bega
     \epsilon_{\bm q}  =  R_{\bm\epsilon,\bm q}  + A T_{\bm\epsilon,\bm q},\quad \epsilon_{\bm q}  = - R_{\bm \epsilon,\bm q} + B T_{\bm\epsilon,\bm q} ,\\
     A\equiv \sqrt{\varepsilon_2\ov \varepsilon_1},\quad B = {\cos\theta_T\ov \cos\theta_I} = {\sqrt{1- A^{-2} \sin^2\theta_I} \ov \cos\theta_I}
\end{gather}

These can be expressed in the matrix equation 
\be
     \begin{pmatrix}
          1 & A \\
          -1 & B 
     \end{pmatrix}
     \begin{pmatrix}
          R_{\bm\epsilon,\bm q} \\ T_{\bm\epsilon,\bm q}
     \end{pmatrix} =
     \begin{pmatrix}
          1 \\ 1
     \end{pmatrix} \epsilon_{\bm q}. 
\ee
We can now invert this matrix to obtain equations for $R,T$ (and noting that $\epsilon_{\bm q}=1$ by normalization assumption), 
\be
     R_{\bm\epsilon,\bm q} = 1 - {2 A\ov A+  B},\quad T_{\bm\epsilon,\bm q} = {2 \ov A+B}. 
\ee
Now use the above equations in conjunction with the fact that 
\be
     \epsilon^z_{\bm q} = \sin\theta_I ,\quad R_{\bm\epsilon,\bm q}^z= -R_{\bm\epsilon,\bm q} \sin\theta_I ,\quad T_{\bm\epsilon,\bm q}^z = T_{\bm\epsilon,\bm q} \sin\theta_T,
\ee
we find
\bega
     \eta_{\bm\epsilon,\bm q} =  {2(1-A^{-2}) \ov 1+A^{-1} B} \sin\theta_I. 
\end{gather}

Comparing with the results of~\cite{Chen2022hydrogel}, the absorptance can be expressed as

\be
\gamma_{\rm Feibleman} (\bm q,\bm \epsilon_{\bm q})= {\omega_{\bm q}  |\eta_{\bm\epsilon,\bm q}|^2 \ov c \cos\theta_I} \left(d_{\perp i} + {d_{\parallel i}\ov  B^2 \tan^2 \theta_I  } \right),
\ee
where $d_{\perp i}$ ($d_{\parallel i}$) is the imaginary part of the perpendicular (parallel) Feibelman parameter. Our model predicts an absorptance of 
\be
     \gamma(\bm q,\bm \epsilon_{\bm q}) = {4\sigma \mu^2 \ov 3 \kappa_0^2 \ell^2 \varepsilon_0 c\hbar}  {|\eta_{\bm \epsilon,\bm q}|^2 \ov \cos\theta_I} \sum_{n=1}^\infty n^{-a} W_n D_n F(1-\Delta E n^a /\hbar\omega_{\bm q}),\quad \kappa_0 \equiv {\sqrt{2 M \Delta E}\ov \hbar}.   
\ee
We therefore find that our model predicts 
\be
     d_{\perp i} = {4\sigma\mu_z^2 \ov 3 \kappa_0^2 \ell^2 \varepsilon_0 \hbar } {F(1-E_B/\hbar\omega_{\bm q})\ov \omega_{\bm q}} \sum_{n=1}^\infty n^{-a} W_n D_n ,\quad d_{\parallel i }= 0
\ee

Based on the parameters determined in main content with different mode function, the corresponding $d_{\perp i}$ ranges from 4.6 $\textup{~\AA}$, 17.5 $\textup{~\AA}$ to 28.6 $\textup{~\AA}$, close to the concluded results in previous literature around 18.8 $\textup{~\AA}$.



\section{Degeneracy suppression factor}

In the long-wavelength limit, it is well-known that a photon can only excite a single low-lying energy-state of the system. At smaller wavelengths, or when there are sharp gradients in the photon mode function, higher-energy states can be excited by a single photon. In the case of a water cluster, such higher-energy modes correspond to exciting multiple rotational states, which is possible as there are strong interactions between molecules. The probability of exciting higher-energy modes, however, is suppressed by a ratio of length-scales, which vanishes in the standard long-wavelength limit. We will consider a very simple example here to illustrate this suppression factor. In particular, we (very crudely) model the internal degrees of freedom of the cluster as a one-dimensional simple harmonic oscillator (SHO) with coordinates $\xi$. The center of mass of the cluster is given by $\bm x=(x,y,z)$; the first-quantized Hamiltonian is then
\be
     H = -{\hbar^2\ov 2M} \pp_{\bm x}^2 - {\hbar^2\ov 2m} \p_\xi^2 + V_0 \delta^{(3)}(\bm x) + \ha m \Omega^2 (z-\xi)^2. 
\ee
Supposing that $m\ll M$, we may solve the eigenvalue problem using the Born-Oppenheimer approximation, where the full wave function is given by $\Psi(\bm x,\xi) = \psi(\bm x) \lam(z-\xi)$. Here $\psi(\bm x)$ is an eigenstate of the delta-function-potential Hamiltonian and $\lam(z-\xi)$ is an eigenfunction of the SHO with mass $m$, frequency $\Omega$, and equilibrium point at $z-\xi=0$. Define the SHO ladder operators by
\be
     c^\pm = {1\ov \sqrt 2} \left( {1\ov l_{\rm SHO}}(\xi-z) \pm l_{\rm SHO} \p_{\xi-z} \right),
\ee
where the characteristic quantum length scale is 
\be
     l_{\rm SHO} = \sqrt{\hbar\ov m\Omega}. 
\ee
When we consider the vibrational, rotational or collective motion of the molecules, we should identify $l_{\rm SHO}$ with $l_{\rm vib}$, $l_{\rm rot}$ or $l_{\rm coll}$, respectively. 

Suppose the cluster begins in the SHO ground state and the delta-function bound state $\ket{\psi_B,0}$ and, upon vaporization by the photon, enters the $\nu^{\rm th}$ SHO excited state and the scattering delta-function state of momentum $\hbar \bm k$, that is $\ket{\psi_{\bm k},\nu}$. By Fermi's golden rule, the rate of this vaporization process is give by
\be
     \Gamma^{(\nu)} \sim | \bra{\psi_{\bm k} ,\nu} H^{(1)} \ket{\psi_B,0}|^2. 
\ee
Noting that $H^{(1)}\sim f^z_{\bm \epsilon ,\bm q}$, $\bra{\psi_{\bm k} ,0}\sim \bra{\psi_{\bm k} ,0} c^\nu$, and $[c, f^z_{\bm \epsilon,\bm q}]\sim l_{\rm SHO} \p_z f^z_{\bm \epsilon,\bm q} \sim (l_{\rm SHO} /\ell) f^z_{\bm \epsilon,\bm q}$ we find 
\be
     \Gamma^{(\nu)} \sim  (l_{\rm SHO}/\ell)^{2\nu} \Gamma^{(0)}, \quad \Gamma^{(0)}\sim | \bra{\psi_{\bm k} \lam_0} H^{(1)} \ket{\psi_B\lam_0}|^2,
\ee
where $\Gamma^{(0)}$ is the rate at which a cluster is vaporized and no internal degrees of freedom are excited. We therefore see that the rate of exciting each additional internal quantum is suppressed by a factor of $\chi^{-1}$, where 
\be
     \chi = \left({\ell\ov l_{\rm SHO} }\right)^2. 
\ee

\end{document}